\documentclass[
    ,final            
  ]
  {aipproc}

\layoutstyle{6x9}

\begin{document}

\title{Introducing a Hybrid Method of Radiative Transfer in Smoothed Particle Hydrodynamics}

\classification{95.30Jx, 95.30.Lz, 95.75.Pq}
\keywords      {radiative transfer, hydrodynamics, computational methods, star formation, planet formation}

\author{Duncan Forgan}{
  address={Institute for Astronomy, Royal Observatory Edinburgh}
}

\author{Ken Rice}{
  address={Institute for Astronomy, Royal Observatory Edinburgh}
}

\author{Dimitris Stamatellos}{
  address={School of Physics and Astronomy, Cardiff University}  
}

\author{Anthony Whitworth}{
  address={School of Physics and Astronomy, Cardiff University}
}

\begin{abstract}

\noindent We present a new method of incorporating radiative transfer into Smoothed Particle Hydrodynamics (SPH).  There have been many recent attempts at radiative transfer in SPH \cite{Stam_2005_1,Stam_2005_2,Mayer,WB_1}, however these are becoming increasingly complex, with some methods requiring the photosphere to be mapped (which is often of non-trivial geometric shape), and extra conditions to be applied there (matching atmospheres as in \cite{Mejia_4}, or specifying cooling at the photosphere as in \cite{Mayer}).  The method of identifying the photosphere is usually a significant addition to the total simulation runtime, and often requires extra free parameters, the changing of which will affect the final results. Our method is not affected by such concerns, as the photosphere is constructed implicitly by the algorithm without the need for extra free parameters.  The algorithm used is a synergy of two current formalisms for radiative effects: a) the polytropic cooling formalism proposed by \cite{Stam_2007}, and b) flux-limited diffusion, used by many authors to simulate radiation transport in the optically thick regime (e.g. \cite{Mayer}).  We present several tests of this method:

\begin{itemize}
\item The evolution of a \(0.07\, M_{\odot}\) protoplanetary disc around a \(0.5\,M_{\odot}\) star (\cite{Mejia_1,Mejia_2,Mejia_3,Mejia_4})
\item The collapse of a non-rotating \(1\,M_{\odot}\) molecular cloud (\cite{Masunaga_1,Stam_2007}) 
\item The thermal relaxation of temperature fluctuations in an static homogeneous sphere (\cite{Masu_98,Spiegel,Stam_2007})
\end{itemize}
\end{abstract}

\maketitle


\section{Methods}

\subsection{The Equation of State}

\noindent SPH alone only evolves the density and internal energy of each particle: radiative transfer requires extra variables such as temperature and opacity.  Therefore, some kind of prescription is required to calculate all required variables using only the density and internal energy.  The prescription used in this work is split into two expressions: the \emph{equation of state} which calculates temperatures and mean molecular weights, and the \emph{opacity law} which calculates opacities.  The equation of state is the same as used by \cite{Stam_2007}: it accounts for the vibrational and rotational energy states of hydrogen and helium, as well as their various dissociation and ionisation states.

\subsection{Polytropic Cooling}

\noindent This approximation uses an SPH particle's density \(\rho_i\), temperature \(T_i\), and gravitational potential \(\psi_i\) to estimate a mean optical depth for the particle \cite{Stam_2007}.  This relies on calculating a mass averaged column density, and a mass averaged opacity.  This averaging process allows us to account for the particle's surroundings: hot particles may be surrounded by a cooler, denser environment, increasing the effective opacity.  The formalism guarantees optimum cooling at the photosphere - however, this formalism cannot model detailed energy exchange between particles.

\subsection{Flux Limited Diffusion}

\noindent This approximation handles the exchange of energy between pairs of neighbouring particles, using the diffusion approximation.  This approximation is only valid in the optically thick regime, so a flux limiter is used to allow the approximation to be expanded to less optically thick situations.  Energy flows along temperature gradients in accordance with the laws of thermodynamics.  However, energy exchange terms are \emph{pairwise}: hence there is no energy loss from the system.

\subsection{The Hybrid Method}

\noindent Comparing the limitations of the above two methods, it should be clear that a union of these two procedures should be complementary: polytropic cooling handles the important energy loss from the system (which flux-limited diffusion cannot), and flux-limited diffusion handles the detailed exchange of heat between neighbouring fluid elements (which polytropic cooling cannot).  Indeed, polytropic cooling's inability to model the detailed exchange of heat between neighbouring fluid elements - and flux-limited diffusion's inability to model energy loss - allow the two methods to work together correctly, modelling all aspects of the system's energy budget without encroaching on each other.

\section{Tests}

\subsection{The Evolution of a Protoplanetary Disc}

\begin{figure*}
$
\begin{array}{cc}
\includegraphics[scale = 0.4]{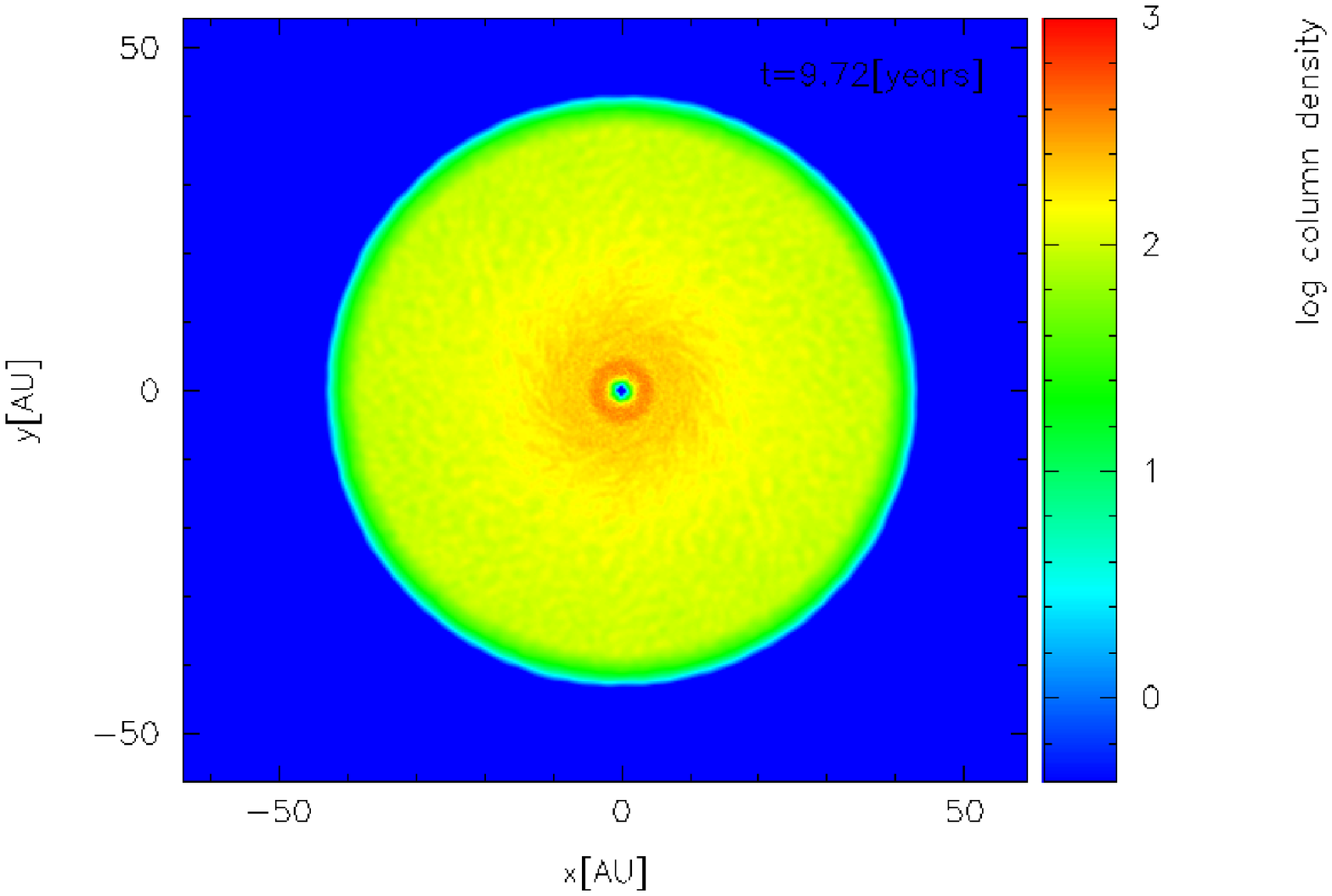} &
\includegraphics[scale = 0.4]{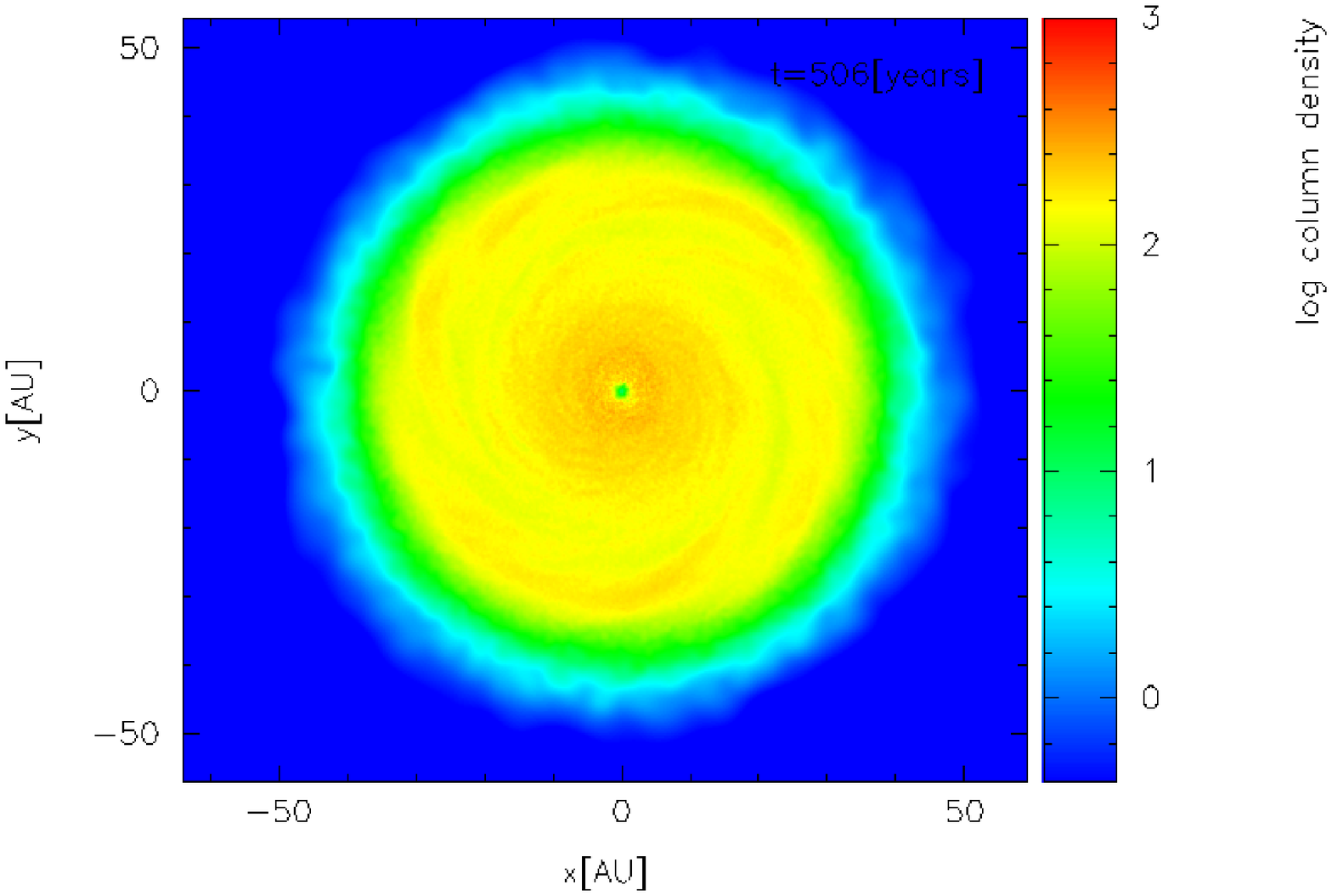} \\ 
\includegraphics[scale = 0.4]{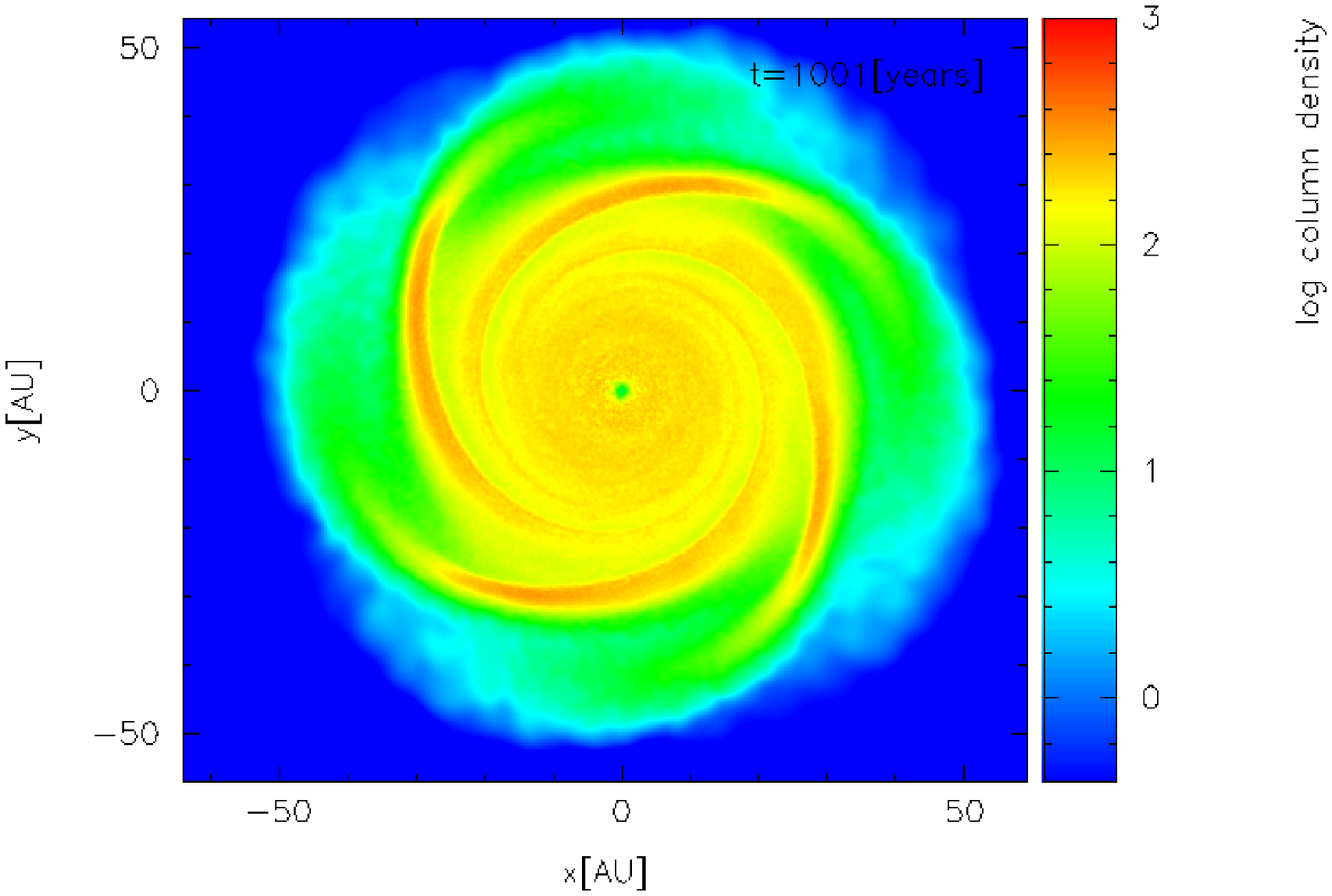} &
\includegraphics[scale = 0.4]{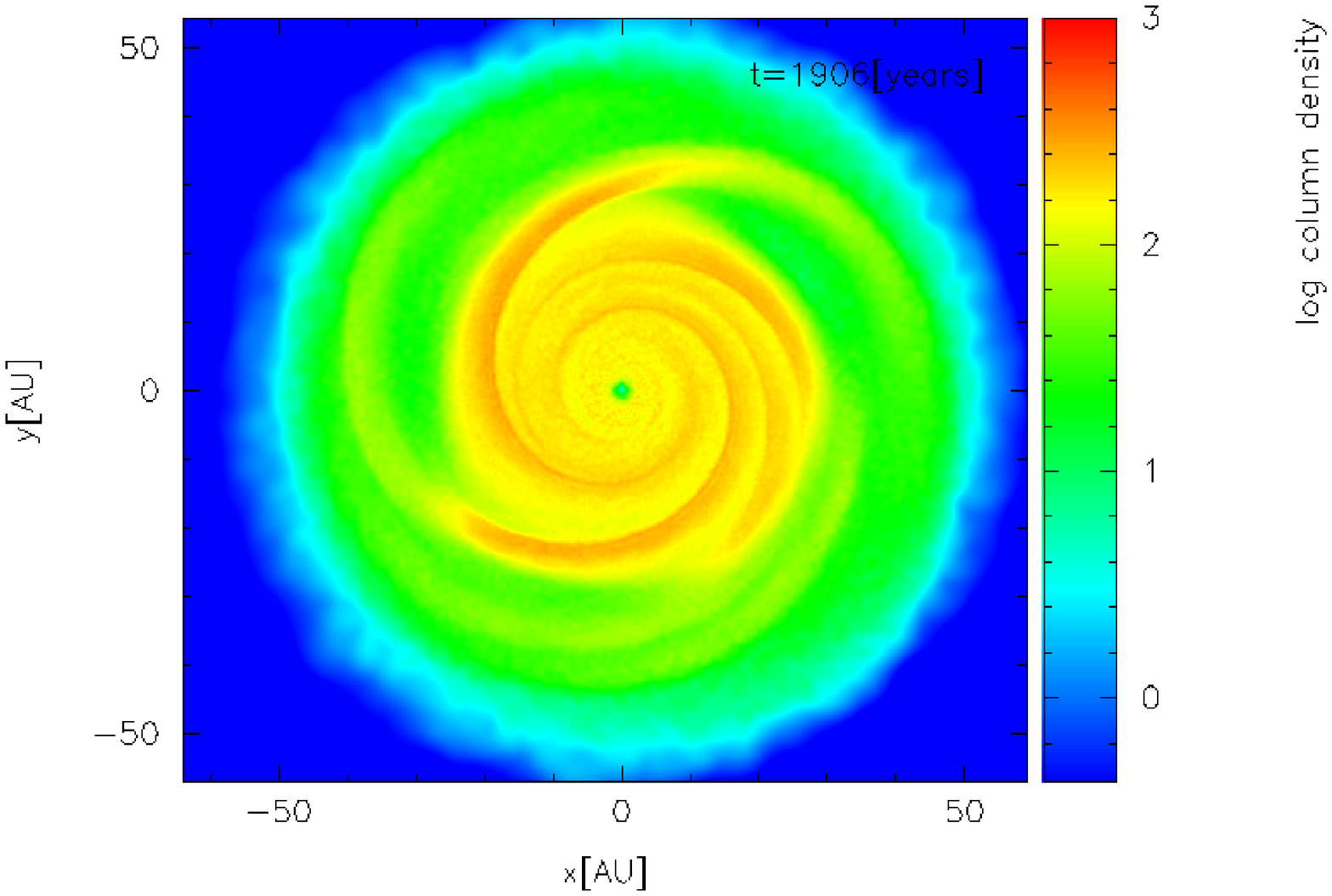} \\ 
\end{array}$
\caption{Density evolution of the protoplanetary disc \label{fig:Mejia_discs}.}

\end{figure*}

\noindent The evolution of a \(0.07\, M_{\odot}\) protoplanetary disc around a \(0.5\,M_{\odot}\) star can be seen in \textbf{Figure \ref{fig:Mejia_discs}}.  The density structure is consistent with the work of Mej\'{i}a, Boley et al \cite{Mejia_1,Mejia_2,Mejia_3,Mejia_4}.  The Toomre Q parameter is at a minimum where the cooling is at a peak, identifying the importance of cooling as a criterion for disc instability.

\subsection{The Collapse of a Molecular Cloud}

\begin{figure*}
$\begin{array}{cc}
\includegraphics[scale=0.5]{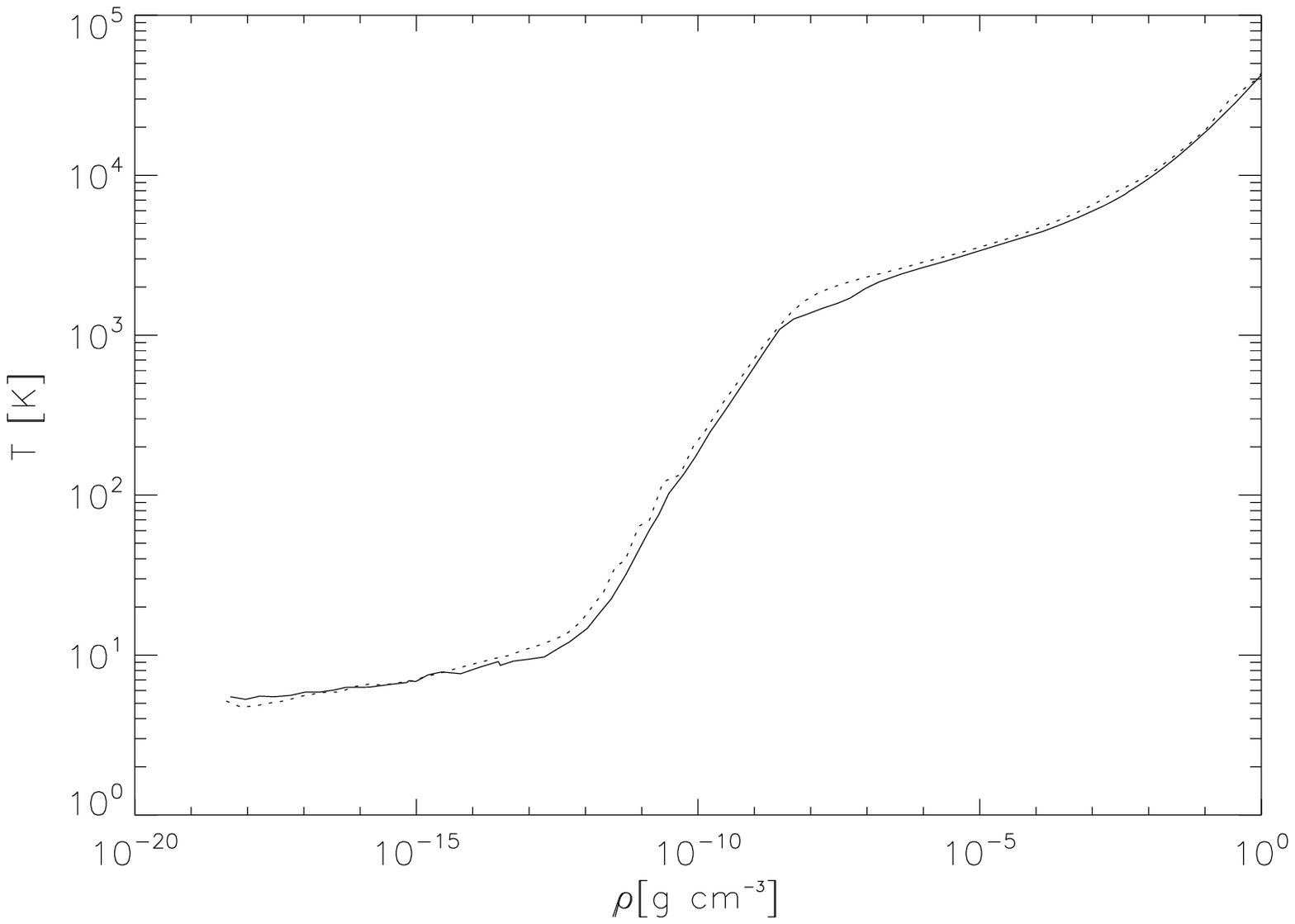} &
\includegraphics[scale=0.5]{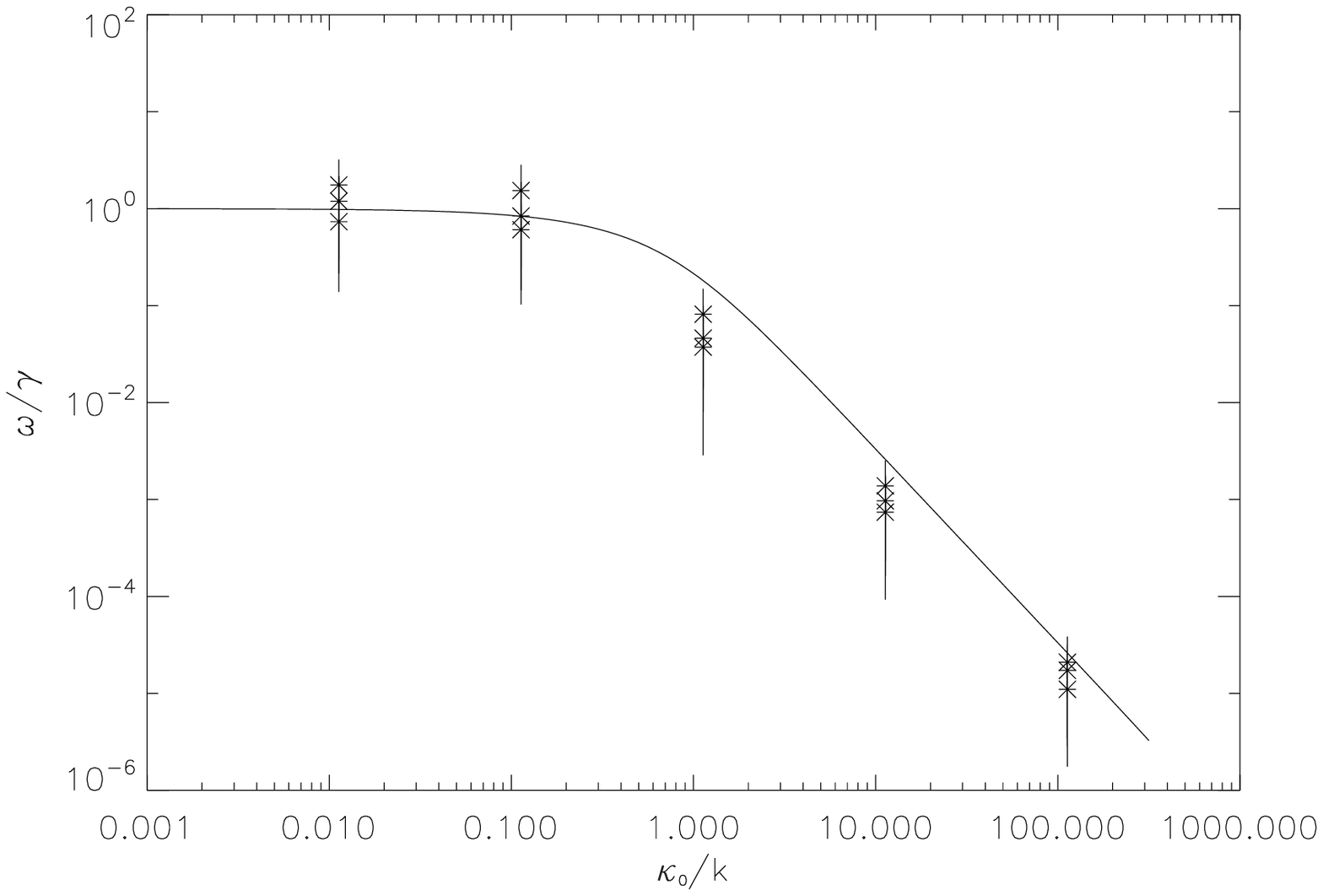} \\
\end{array}$
\caption{Left: Evolution of the central temperature as a function of central density in the collapsing cloud. The dotted line shows the same results for using polytropic cooling only.  Right: The dispersion relation \(\omega\) for the Spiegel Test. \label{fig:sp_omega}}
\end{figure*}

\noindent The collapse of a non-rotating molecular cloud was then simulated.  The spherical, uniform density cloud contains \(1\,M_{\odot}\) of material (populated by \(5 \times 10^5\) SPH particles), and has a radius of \(10^4\) AU (which gives a density of \(\rho_0 = 1.41 \times 10^{-19} \, \mathrm{g\, cm^{-3}}\)).  These conditions were initially investigated by \cite{Masunaga_1} by solving the full radiative transfer in 3D (with the hydrodynamics solved in 1D), and were revisited by \cite{Stam_2007}.  As can be seen from \textbf{Figure \ref{fig:sp_omega}} (left panel), the initial isothermal phase (at low densities) lasts longer for the hybrid method, as extra cooling can occur due to the temperature gradients in the core.

\subsection{The Spiegel Test}

\noindent To study the method's time dependence, the relaxation rate of temperature fluctuations in a static sphere was simulated for a series of clouds of different optical depth.  The lines in each figure show the analytic solution for the relaxation rate \(\omega\) as a function of cloud opacity (divided by the wavenumber, k, of the fluctuation) \cite{Masu_98,Spiegel,Stam_2007}.  As can be seen, the hybrid method approximates the analytic solution well.

\section{Conclusions}

\noindent We have created a new radiative transfer algorithm by merging two currently existing methods.  By doing so, the algorithm can model in detail frequency-averaged radiative transfer.  Several tests of the algorithm against both analytic and numerical benchmarks have shown it to be successful in capturing the necessary physics.

\begin{theacknowledgments}
 
\noindent The authors would like to acknowledge D. Price for the use of SPLASH \citep{Price}.  All simulations were performed using the supa64 machine at the University of St. Andrews.  DS and AW gratefully acknowledge the support of an STFC rolling grant (PP/E000967/1) and a Marie Curie Research Training Network (MRTN-CT2006-035890).

\end{theacknowledgments}





\end{document}